\documentclass[journal]{IEEEtran}
\usepackage{CJK}
\usepackage{amssymb}
\usepackage{amsmath}
\usepackage{graphicx, subfigure}
\usepackage{url, cite}
\usepackage{verbatim}
\usepackage{booktabs}
\usepackage{multirow}
\usepackage{bigstrut}
\usepackage{booktabs}

\usepackage{amsmath,cite,graphicx,times,subfigure,url,verbatim}
\usepackage{algorithmic}
\usepackage{algorithm}

\newtheorem{subsec:coding}{subsec:coding}

\usepackage{color}

\begin{document}

\title{Label-less Learning for Traffic Control in an Edge Network}

\author{
Min~Chen,~Yixue~Hao,~Kai~Lin,~Zhiyong~Yuan,~Long~Hu

\thanks{M. Chen, Y. Hao and L. Hu are with Huazhong University of Science and Technology, China. Email: minchen2012@hust.edu.cn}
\thanks{K. Lin is with Dalian University of Technology, China. Email: link@dlut.edu.cn}
\thanks{Z. Yuan is with Wuhan University, China. Email: zhiyongyuan@whu.edu.cn}
\thanks{Long Hu is the corresponding author}
}

\markboth{Under review:, VOL. XX, NO. YY, MONTH 20XX}{}

\maketitle

\begin{abstract}
With the development of intelligent applications (e.g., self-driving, real-time emotion recognition, etc), there are higher requirements for the cloud intelligence. However, cloud intelligence depends on the multi-modal data collected by user equipments (UEs). Due to the limited capacity of network bandwidth, offloading all data generated from the UEs to the remote cloud is impractical. Thus, in this article, we consider the challenging issue of achieving a certain level of cloud intelligence while reducing network traffic. In order to solve this problem, we design a traffic control algorithm based on label-less learning on the edge cloud, which is dubbed as LLTC. By the use of the limited computing and storage resources at edge cloud, LLTC evaluates the value of data, which will be offloaded. Specifically, we first give a statement of the problem and the system architecture. Then, we design the LLTC algorithm in detail. Finally, we set up the system testbed. Experimental results show that the proposed LLTC can guarantee the required cloud intelligence while minimizing the amount of data transmission.
\end{abstract}


\section{Introduction}

With the recent development of Internet of Things (IoT) and cloud computing, IoT has facilitated the new generation of intelligent applications (such as autonomous driving, virtual reality, and augmented reality)~\cite{1,2}. In addition to requiring a lot of data transmissions and computation to acquire intelligence, these new applications also have higher sensitivity to latency. For example, in autonomous driving, it is necessary to enhance the accuracy of data analysis, and make decisions quickly~\cite{3}. As a result, these services are smarter, faster, and interact with users more frequently. Thus, these new services have a need for network bandwidth and the ever-increasing cloud intelligence.

With the development of Artificial Intelligence (AI), deep learning has achieved excellent performance in extracting the features of unstructured data, such as image and speech~\cite{4,5}. Thus, cloud intelligence based on deep learning has made great progress. However, deep learning relies on a large number of data sets, and real-time cloud intelligence depends on the continuous provisioning of application-aware data generated by terminal devices. Fortunately, with the rapid growth of user equipments (UEs), the data collected by the UEs exhibit the exponential growth, which is able to provide the foundation for the requirements in order to achieve cloud intelligence via real-time analysis~\cite{6}. However, due to the limited capacity of network bandwidth, offloading all data to the cloud results in the problem of high energy consumption, as well as it to fail in meeting application demand of low latency~\cite{7}. Therefore, offloading the mass data to the cloud is impractical. How to reduce the data amount being offloaded to the cloud is a challenge problem.

Communication latency can be reduced and the data amount transmitted to the cloud can be decreased with the development of edge computing~\cite{8,9}. This can be done by deploying an edge server with storage and computing capacity on the network edge (such as a router and small cell base station), because the edge server is closer to the user and only needs one hop in most cases. However, due to limited computing and storage capacity of the edge server, it is difficult for the existing edge computing mode to meet the needs of intelligent applications that require real-time processing~\cite{10}. Thus, how to reduce the amount of data as far as possible without weakening system intelligence is a challenging problem that needs to be solved in order to utilize edge computing for intelligent applications.

\begin{figure*}
\centerline{\includegraphics[width=6.0in]{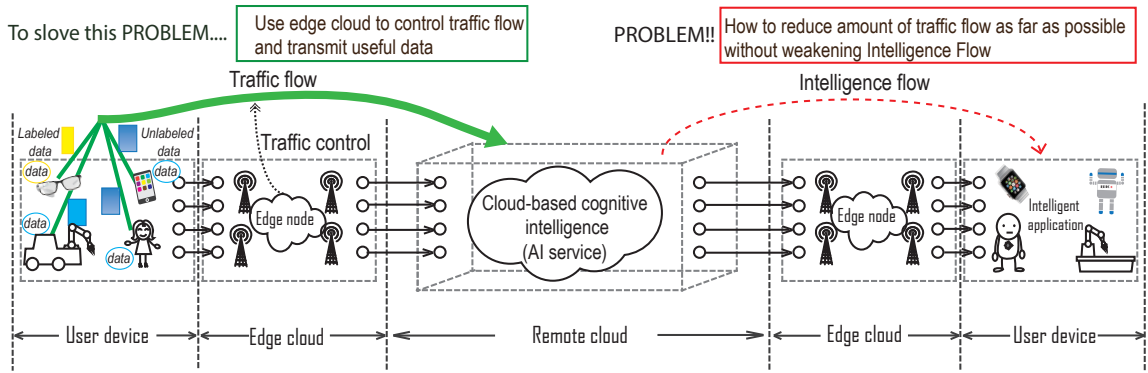}}
\caption{An illustration of traffic control in edge cloud.}
\label{fig01}
\end{figure*}

In this article, to overcome the challenges described above, we consider the problem of how to control the network traffic flow by the limited computing and communication resources of the edge cloud. Specifically, we consider the system architecture involved in AI-based applications, the edge cloud, and the remote cloud. In view of a lot of the unlabeled data collected by the UEs, we propose lablel-less learning to to filter unlabeled data. Rather than offloading to the cloud directly, this article design a Label-less Learning based Traffic Control (LLTC) algorithm
on the edge cloud to reduce the data offloading while maintaining cloud intelligence to a certain level. In LLTC, after labeling and selecting, the data are offloaded to the cloud, so as to ensure the application-aware intelligence in the remote cloud while reducing network traffic transmissions.

Specifically, we describe our idea by using the example of real-time emotion recognition and interaction, as shown in Fig.~\ref{fig01}. The wearable device collects the real-time emotion data (such as the facial expression and speech)~\cite{11,12}. The data includes a small amount of labeled data and a large amount of unlabeled data. Various data flows aggregate into a traffic flow, and those data are delivered to the edge cloud. The raw data is analyzed and selected by label-less learning in the edge cloud. Only those data that can enhance the accuracy rate of the emotion model are offloaded to the remote cloud. In the remote cloud, intelligence (i.e., the accuracy of emotion detection) is gained after real-time data analysis and training through deep learning. Finally, the real-time emotion detection is enabled for the emotion-aware applications. Thus, we utilize a small amount data to realize a stronger intelligence of the cloud through label-less learning data control and reduce network resource consumption.


The main contributions of this article are list as follows:
\begin{itemize}
\item In view of the real-time and intelligent requirements of applications, we propose the problem of how to minimize traffic flow transmission while ensuring cloud intelligence.. To the best of our knowledge, this problem is being researched for the first time.
\item In view of above problem, we design a LLTC algorithm on the edge cloud. After the labeling and selecting of unlabeled data, the data offloading to the remote cloud is reduced and the cloud intelligence is enhanced.
\item We set up a testbed based on emotion recognition to verify our LLTC algorithm. The experimental results indicated that the algorithm proposed by this article reduced data transmission on the basis of ensuring the accuracy rate of emotion detection.
\end{itemize}

The remainder of this article is organized as follows: In Section \ref{sec:arch}, problem statement and system architecture is introduced. Label-less learning based traffic control in edge cloud is given in Section ~\ref{sec:tech}. In Section \ref{sec:design}, the system testbed and experimental results are presented. Finally, conclusions are given in Section \ref{sec.conclusion}.

\section{Problem Statement and System Overview}   \label{sec:arch}

In this section, we first introduce the problem statement, as shown in Fig.~\ref{fig02}. Then, we provide the system overview.

\subsection{Problem Statement}

\begin{figure*}
\centerline{\includegraphics[width=6.2in]{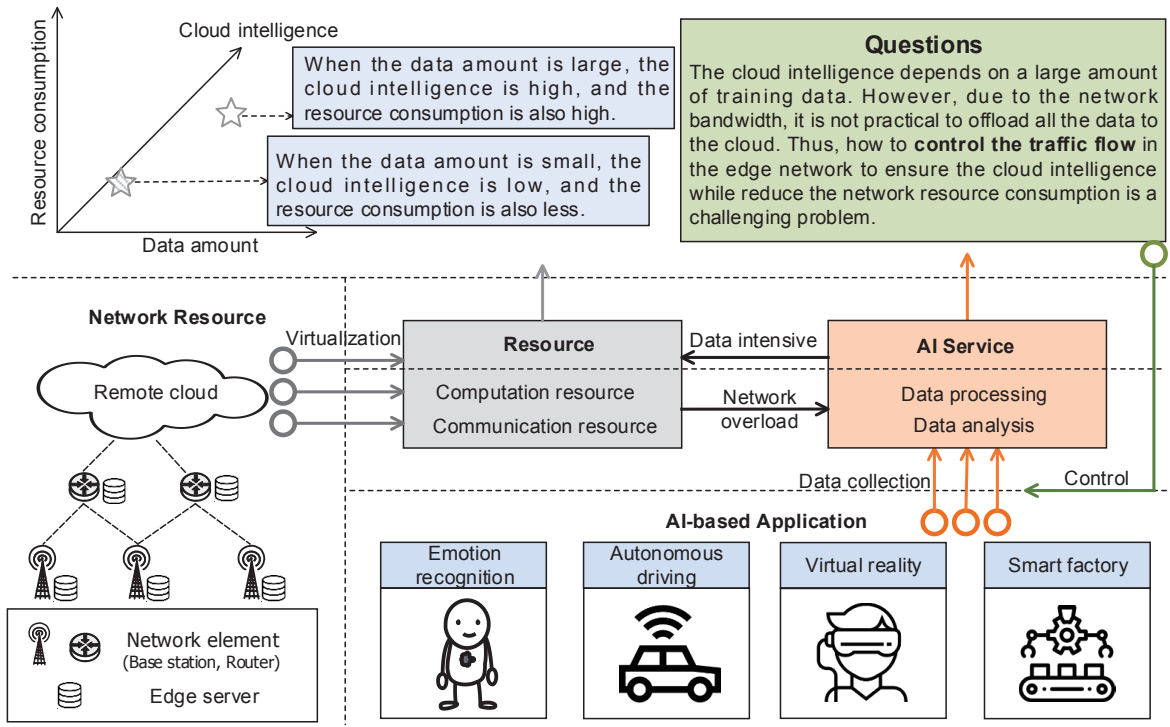}}
\caption{The trade-off among cloud intelligence, data amount and resource consumption.}
\label{fig02}
\end{figure*}

With the development of intelligent applications, more and more services are based on AI, as shown in Fig.~\ref{fig02}. These AI-based applications include real-time emotion recognition (requiring real-time emotion data collection and recognition), autonomous driving (requiring real-time road condition data collection and analysis), virtual reality (requiring real-time image data collection and analysis), and smart factories (requiring real-time device data and analysis). These applications need an effective combination of data collection and analysis to perform. However, the traditional cloud architecture cannot support these real-time services based on AI.

Our aim is to minimize the amount of data traffic flowing while maintaining the cloud intelligence. We can provide the problem statement using the example of emotion recognition. For the real-time emotion detection, we should first consider the accuracy of emotion recognition. The terminal-based emotion collection device can collect multidimensional and multimodal data, such as collecting the physiological, speech, and facial emotion data of the user, etc. If offloading all the collected data to the cloud for emotion recognition model training, a high prediction model can be obtained to guarantee the accuracy rate of emotion recognition. However, such a mode will be restricted by the limited capacity of network bandwidth. This results in a higher communication latency and accordingly fails to meet real-time requirement of application. Although multimodal emotion data recognition on the edge cloud can reduce the communication latency, the computing capacity of the edge cloud is limited and results in higher computation latency. Thus, during the process of emotion recognition, ensuring the accuracy rate of emotion recognition, while also reducing the amount of data transmission to the cloud is a challenging problem.

It should be noted that the data collected by the UEs includes three types of data: a small amount of labeled data, a large amount of unlabeled data, and noise data. In this article, we assume the collected data is multimodal data. The unlabeled data and noise data include huge amounts of unnecessary data. Offloading these unnecessary data to the cloud has two disadvantages: (i) offloading increases the training complexity of machine learning in the cloud, without having a large influence on cloud intelligence; and (ii), offloading these unnecessary data to the cloud results in higher communication latency and causes an unpredictable quality of service degradation, which causes it to fail to meet the requirements of real-time service.

Furthermore, the existing deep learning algorithm depends on having a large amount of labeled data. Thus, the unlabeled data should be labeled. Furthermore, to realize the traffic control problem, the main idea is to filter the unnecessary data in the edge cloud, and reduce the amount of data transmission while maintaining cloud intelligence. Therefore, in this article, we convert the traffic flow control problem into a labeling and selecting problem of unlabeled data in the edge cloud. We will introduce it detain in Section~\ref{sec:tech}

\subsection{System Overview}

The system architecture includes three layers: the AI-based application, edge cloud, and remote cloud. The application layer is the service requester and data collector. The edge cloud is composed of many edge nodes with computing and storage capacity. These nodes can be the base station, gateway, and router, etc. The edge node is closer to the UEs, so the communication latency and resource consumption is lower during the data transmission. However, the edge computing node has limited computing and storage capacity and it is difficult to finish the comparatively complicated computing task, such as a deep learning model in need of multimodal data. Thus, it only can process the tasks with smaller computational complexity and give the real-time feedback. The remote cloud is composed of a data center, as the source of the intelligent application's wisdom. In general cases, the remote cloud can learn the knowledge and give feedback to the application users by the large amount of training data and a deep learning algorithm.

We design a LLTC algorithm. The details are as follows: First, the AI-based application offloads the collected unlabeled multimodal data to the edge cloud through a cellular network or WiFi. On the edge cloud, based on label-less learning, the unlabeled data is labeled and selected. The communication delay is much lower because the edge cloud is close to UE. Secondly, the edge cloud offloads the preliminary selected data to the remote cloud for processing. Through the preliminary selecting of unlabeled data on edge cloud, the data amount is greatly reduced and the data includes only the most useful information. Thus, the intelligence of the remote cloud is also increasing.

\begin{figure*}
\centering
\includegraphics[width=6.0in]{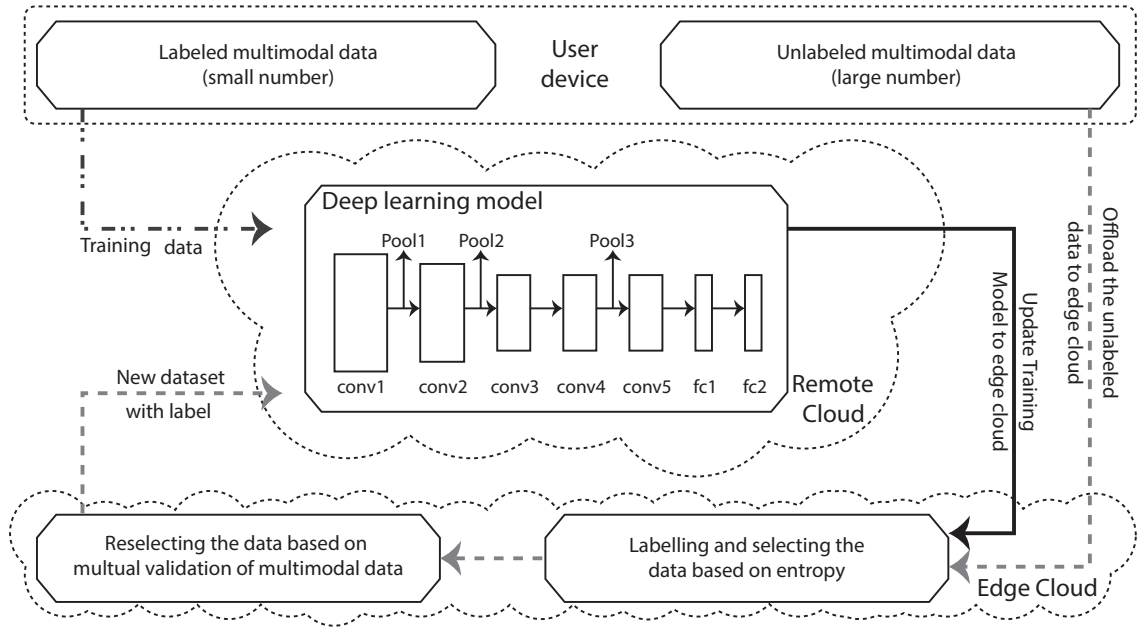}
\caption{Label-less learning based traffic control in edge cloud.}
\label{fig03}
\end{figure*}

Thus, the LLTC algorithm realizes the traffic control. Based on the traffic control, the cloud intelligence only need little data and the resource utilization rate is enhanced. The details as follows:
\begin{itemize}
\item Cloud intelligence: The remote cloud utilizes the data offloaded from the edge cloud, trains the data based on the deep learning algorithm, and obtains the highly intelligent model. At the same time, this model is pushed to the edge cloud. The edge cloud first labels and selects the unlabeled data based on label-less learning and this model, then offloads the most useful data to the cloud. Therefore, the system forms the closed loop of data collection, data analysis, and data control, and promotes the system intelligence.
\item Virtualization: We utilize the virtualization techniques to virtualize the communication and computing resources of edge cloud and utilize network slicing technology to realize the isolation of different applications and processing flows, such as processing autonomous driving and a smart factory through different network slicing, so as to meet the latency requirements of AI-based application
\item Resource utilization rate: Based on the traffic control algorithm, we make the most of the limited computing and storage resources of edge cloud and the computing resources of the cloud. Through the mutual cooperation of the edge cloud, remote cloud, and AI, the system dynamically utilizes the computing and communication resources to make a higher resource utilization rate.
\end{itemize}

\section{Label-less Learning based Traffic Control} \label{sec:tech}

In this section, we introduce LLTC in edge cloud. We first give an overview of label-less learning. Then, we propose label-less learning based on entropy. Thirdly, we introduce multimodal data control based label-less learning.

\subsection{An Overview Label-less Learning}

With the advances on networking technology, the unlabeled data acquisition has become more and more convenient. Thus, how to label the collected massive unlabeled data is a challenging issue. Existing work mainly adopt positive and unlabeled learning (i.e., PU learning), self-training and co-training. PU learning refers to the training dataset consisting of a small amount of labeled positive data and a large amount of unlabeled data. Since PU learning only contains two classes, so it cannot be directly applied to the unlabeled data with multiple classes. A self-training is a common algorithm for inductive semi-supervised learning and can label the unlabeled data automatically. However, the self-training method generally fails to judge whether the automatic label is correct. After adding the wrong automatically-labeled data to the training set, the error accumulation is caused. The co-training is a common unsupervised learning from the multi-view perspective. It can validate each other among multimodal data, and can enhance the confidence level of the new labeled data to some extent. However, the selecting of unlabeled data has the limitation.

The label-less learning refers to a kind of learning method of model recognition on many unlabeled multimodal data. It refers to how to label multimodal unlabeled data, and how to add these unlabeled data into training model. The target of label-less learning is to improve the accuracy of classification through utilizing a mass of unlabeled data. Thus, through label-less learning, we can realize automatic labeling for unlabeled data. This not only reduces the dependence on labeled data, but also reduces the waste of human and material resources required for labelling data. Next, we will give the specific process of label-less learning.

The label-less learning can be divided into following four steps:
\begin{itemize}
\item We use small amount of labeled data as a set for training the model to provide the system with initial intelligence. In this paper, we use the deep convolution neural network (i.e., AlexNet) algorithm~\cite{15} to train the model.
\item We use the above mentioned training model to label the unlabeled data, select a portion of the new labeled data, and add it into training set.
\item Based on the mutual verification of multimodal data, we make an additional selecting for the newly added data.
\item We train the model again for the newly added labeled data to make the system more intelligent.
\end{itemize}
In the above-mentioned steps, we can see following problems associated with label-less learning: (i) How to label the unlabeled data and add it into the training set; (ii) For multimodal data, how to realize mutual verification of the multimodal data set. In the next, we will introduce the label-less learning in detail.

\subsection{Label-less Learning based on Entropy }

We assume the labeled data sets are $\mathbf{x^{l}}=(\mathbf{x_{1}^{1}}, \mathbf{x_{2}^{l}}, \cdots, \mathbf{x_{n}^{l}})$, where $n$ represents the number of data sets with labels. The labels of data sets are $\mathbf{y}=(y_{\mathbf{x_{1}^{l}}}, y_{\mathbf{x_{2}^{l}}},\cdots, y_{\mathbf{x_{n}^{l}}})$. The unlabeled data sets are $\mathbf{x^{u}}=( \mathbf{x_{1}^{u}}, \mathbf{x_{2}^{u}} \cdots,\mathbf{x_{m}^{u}})$, where $m$ represents the number of data sets without labels. We suppose that the unlabeled data exceeds the labeled data. Our goal is to label the unlabeled data sets and put those new labeled data into the training sets.

In this article, we adopt the predicting uncertainty to select the new data to be labeled, i.e., only the data with a low uncertainty will be selected through the prediction. We use entropy as a measure to estimate the prediction uncertainty. We assume that the predictive probability of the unlabeled data $\mathbf{x_{i}^{u}}$ is $\mathbf{y_{\mathbf{x_{i}^{u}}}}=\{p_{\mathbf{x_{i}^{u}}}^{1}, p_{\mathbf{x_{i}^{u}}}^{2}, \cdots, p_{\mathbf{x_{i}^{u}}}^{c}\}$, where $p_{\mathbf{x_{i}^{u}}}^{j}$ represents the predicted probability value of classification $j$, and $c$ represents the number of classifications. Then, the entropy of the predicted probability for the unlabeled data is defined by:
\begin{equation}\label{eq:1}
E(\mathbf{y_{\mathbf{x_{i}^{u}}}})=-\sum_{j=1}^{c}p_{\mathbf{x_{i}^{u}}}^{j}\log(p_{\mathbf{x_{i}^{u}}}^{j})
\end{equation}
In \eqref{eq:1}, when the entropy value is relatively small, the newly labeled data have a lower prediction uncertainty. Therefore, entropy can be used as a standard for selecting the unlabeled data to be labeled.

There might be an accumulated training error if the data selected are based on a low entropy value and are used to train the model as a new training data set. This is because during the process of actual model annotation there could be wrongly-made labels which results in increased noise in the data set, and makes the error relatively large.

To overcome these problems, we propose the following strategies: New labeled data will be re-evaluated when they are added to the training set instead of always relying on the data of a low entropy threshold. Specifically, our evaluation algorithm is as follows: First, we denote the newly-added labeled data set based on a low entropy threshold as $\mathbf{z}$. Then, we re-select the new labeled data from $\mathbf{z}$ through $n$ iterations. We denote that the amount of automatically labeled data in each iteration is $k$, and the added data set is $\mathbf{s}$. Thus, we can obtain $|\mathbf{s}|=k$, where $|\cdot|$ represents the number of elements. For the $\mathbf{x_{i}^{u}} \subseteq \mathbf{s}$ and arbitrary $\mathbf{(x_{i}^{u})'} \subseteq (\mathbf{z}-\mathbf{s})$, the added data need to satisfy the following requirement:
\begin{equation}\label{eq:2}
E(\mathbf{y_{\mathbf{x_{i}^{u}}}}) \leq E(\mathbf{y_{\mathbf{(x_{i}^{u})'}}})
\end{equation}
In the process of specific experiments, we added the same number of data sets to each category in every iteration to maintain the balance of the categories. In addition, we assumed the selected data were progressively increasing in the iterations.

\subsection{Multimodal Data Control based Label-less Learning}

The data collected by the UEs are generally multimodal. For example,  speech and facial expressions are two important modalities for emotion recognition. In this article, for simplicity, we consider only two modal of data. However, selecting for the multi-modal unlabeled data is more complicated compared to the single-modal unlabeled data. This is because: (i) the labels given by different modal data may be inconsistent, how to give the label for the multimodal unlabeled data; and (ii) how the new labeled data are selected to be added into the training set.

In view of the above two problems, the details of labeling and selecting of the multi-modal data are as follows: we assume that one modal of data sets are $\mathbf{xf}$, and the other modal of data sets are $\mathbf{xs}$. Thus, we can obtain $\mathbf{x^{l}}=(\mathbf{xf^{l}}, \mathbf{xs^{l}})$, and $\mathbf{x^{u}}=(\mathbf{xf^{u}}, \mathbf{xs^{u}})$. Furthermore, we can obtain the predicted entropy for the two modal of unlabeled data are $E(\mathbf{y_{\mathbf{xf^{u}}}})$ and $E(\mathbf{y_{\mathbf{xs^{u}}}})$, respectively. We chose the entropy with a smaller value to be the label of the unlabeled data when the labels for the two modal are different.
As to how the new labeled data is added to the training set, we adopt minimum joint prediction entropy strategy, of which the calculation process is as follows:
\begin{equation}\label{eq:3}
\bar{E}(\mathbf{y_{\mathbf{x^{u}}}})=\frac{1}{2}E(\mathbf{y_{\mathbf{xf^{u}}}})+\frac{1}{2}E(\mathbf{y_{\mathbf{xs^{u}}}})
\end{equation}
For the selection of the new labeled multi-modal data, we perform the selection based on~\eqref{eq:2}. This strategy not only avoids the prediction conflicts caused by various modal data, but can also increase the correctness of the automatic label data selection. Thus, we give the multimodal data control based label-less learning.

Finally, we give how to apply the label-less learning for traffic control, as shown in Fig.~\ref{fig03}. First, the collected unlabeled multimodal data is offload to the edge cloud through a cellular network or WiFi. On the edge cloud, based on label-less learning, the unlabeled data is labeled and selected. The communication delay is much lower because the edge cloud is close to UEs. Secondly, the edge cloud offloads the preliminary selected data to the remote cloud for processing. Through the preliminary selecting of unlabeled data on edge cloud, the data amount is greatly reduced and the data includes only the most useful information. Thus, the accuracy of emotion detection in the remote cloud is also increasing.

\begin{figure}
\centering
\includegraphics[width=3.6in]{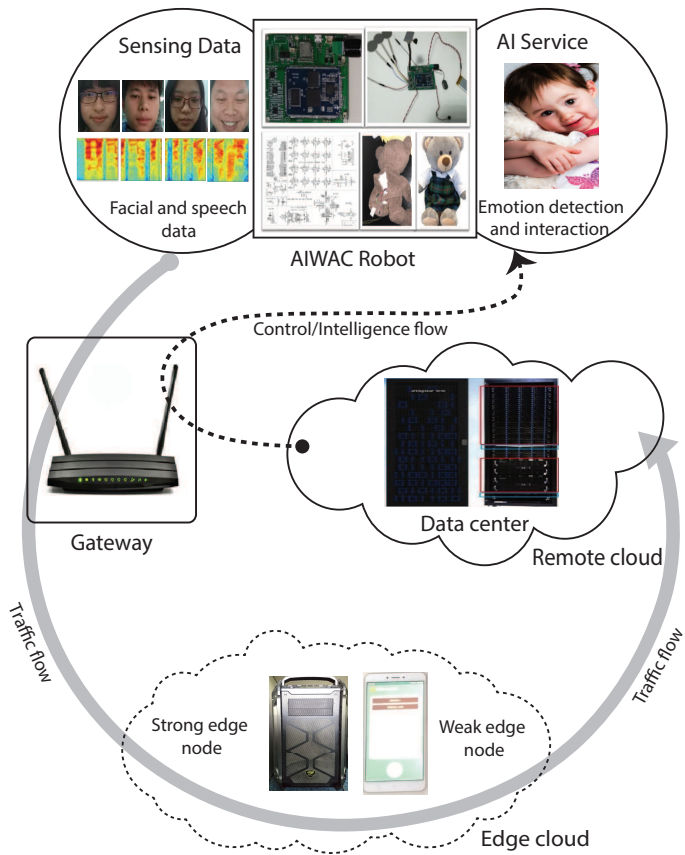}
\caption{System testbed.}
\label{fig04}
\end{figure}

\section{Experimentation Result}
\label{sec:design}

In this section, to verify the LLTC proposed by this article, we set up a system testbed in view of the emotion recognition and interactive application. Then, we measure the LLTC algorithm from two perspectives: the transmitted multimodal data amount and the emotion detection accuracy.

\subsection{System Testbed}

The system testbed we built is shown in Fig.~\ref{fig04}, and involves three parts: the AIWAC (Affective Interaction through Wide Learning and Cognitive Computing) robot~\cite{13}, the edge cloud consisting of smart phones and a small server, and the remote cloud consisting of the data center. In this article, the data collected by the AIWAC robot was transmitted to the edge cloud by WiFi. The AIWAC robot included various sensors such as a camera and microphone, and could collect the data of human facial expressions and speech during the process of emotion recognition.

The edge cloud deployed three algorithms as follows: (i) Deploy the OpenCV~\cite{14} algorithm to realize the human face detection. (ii) Deploy the voiceprint recognition algorithm to detect the speaker. (iii) Deploy the entropy-based label-less learning algorithm with multimodal mutual verification to realize the traffic flow control. The data center deployed the deep convolution neural network (i.e., AlexNet) algorithm~\cite{15}, and conducted the emotion recognition based on the two modal data of image and voice. The basic process was as follows: First, the voice and image data were trained by AlexNet and the voice and image features were obtained. Then, emotion recognition was realized based on the fusion of the feature level.

The specific experimental process was as follows: First, we used labeled image and voice data for training the AlexNet model in a data center, and the trained model was pushed to the edge cloud. The labeled data we used was the enterface05 data set which contained 1,290 videos, and speeches of 43 different speakers, all of which were in English. It included seven basic emotions: anger, disgust, fear, happiness, sadness, surprise, and neutrality. Then, we utilized the real-time data collected by AIWAC to realize the real-time emotion recognition and interactive. Specifically, the unlabeled data (e.g., human face detection and voiceprint recognition) was first preprocessed on the edge cloud. Then, data selecting and labeling was realized in accordance with the model pushed by the data center and the label-less learning algorithm, and the data was offloaded to the data center. In the data center, data was retrained, and the result was fed back to the user, and the model was pushed to the edge cloud.

\subsection{Experimental Results and Analysis}
\begin{figure}
\centering
\includegraphics[width=3.5in]{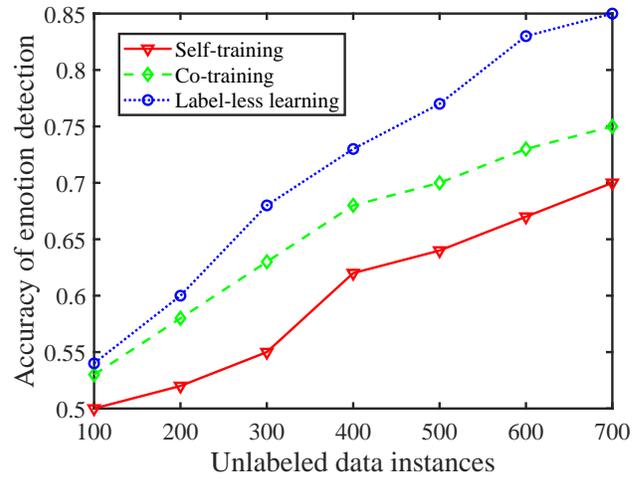}
\caption{Emotion accuracy comparison among self-training, co-training and label-less learning under different number of unlabeled data instances.}
\label{fig11}
\end{figure}

We compare the label-less learning algorithm with self-training and co-training and give the accuracy of emotion recognition under different numbers of unlabeled data. From the Fig.~\ref{fig11}, we can conclude that the enhanced hybrid LEC algorithm proposed in this paper is the best, this is because the proposed enhanced hybrid LEC has two adventages: (i) it can validate emotion data from two modalities of face and image. (ii) it can further filter the unlabeled data of automatic labelling to validate the accuracy of the label data.

\begin{figure}
\centering
\includegraphics[width=3.6in]{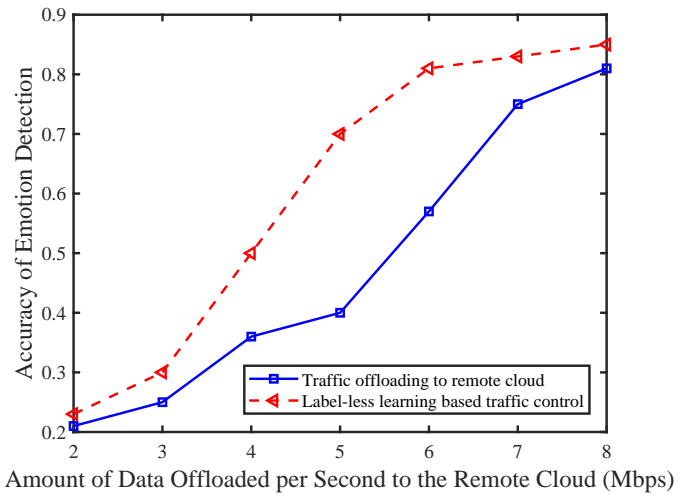}
\caption{The relationship between amount of data offload per second to the remote cloud and accuracy of emotion detection.}
\label{fig06}
\end{figure}

Furthermore, to verify the effectiveness of the LLTC algorithm, we compare the accuracy of emotion detection in two cases of LLTC and traffic offloading to remote cloud at the same data amount transmitted to the cloud. From the Fig.~\ref{fig06}, we can see that the larger the amount of data offloaded to the remote cloud, the higher the accuracy of emotion detection. Furthermore, in the case of LLTC, the accuracy of emotion detection is higher than that of traffic offloading to remote cloud. This is because our algorithm can minimize the data transmission quantity on the premise of maintaining the cloud intelligence.

\section{Conclusion}\label{sec.conclusion}

In this article, we researched the problem of how to maintaining cloud intelligence while reducing network traffic in view of the requirements of AI-based applications on cloud intelligence and real-time performance. We proposed the LLTC algorithm by fully using the computing and communication resources of the edge cloud. This algorithm only offloads the useful data to the cloud by the labeling and selecting of unlabeled data. This traffic control algorithm not only reduced the amount of data transmission, but also had little influence on cloud intelligence. Furthermore, the effectivity of the algorithm proposed by us was verified by building a testbed. In this paper, the LLTC algorithm is still facing challenges especially when dealing with situations requiring context awareness and autonomous decision making under real-time constraints. Thus, in future studies, we will enhance the remote cloud intelligence and reduce the traffic flow through semantic and context prediction on the edge cloud.

\bibliographystyle{IEEEtran}


%
%
%

\end{document}